\documentclass[sigconf,nonacm]{acmart}
\usepackage{color, colortbl}
\usepackage[utf8]{inputenc}
\usepackage{booktabs}
\usepackage{amsmath,amsfonts}
\usepackage{natbib}
\setcitestyle{square, comma, numbers,sort&compress, super}
\usepackage{graphicx}
\usepackage{enumitem}
\usepackage{listings}
\usepackage[linesnumbered,ruled,lined]{algorithm2e}
\usepackage{graphicx}
\usepackage{multirow}
\usepackage{color, soul}
\definecolor{gray}{RGB}{215,215,215}
\sethlcolor{gray}
\usepackage{framed}
\usepackage{hhline}
\usepackage{subcaption}
\usepackage{tcolorbox}
\usepackage{float}
\usepackage{array}
\usepackage{xspace}
\usepackage{varioref}
\usepackage[export]{adjustbox}
\usepackage{tikz}
\def\checkmark{\tikz\fill[scale=0.4](0,.35) -- (.25,0) -- (1,.7) -- (.25,.15) -- cycle;}
\definecolor{mycolor}{rgb}{0.122, 0.435, 0.698}% Rule colour
\definecolor{gray1}{gray}{0.3}

\makeatletter
\def\subsubsection{\@startsection{subsubsection}{3}%
  \z@{.5\linespacing\@plus.7\linespacing}{.1\linespacing}%
    {\normalfont\itshape}}
    \makeatother

    \definecolor{dkgreen}{rgb}{0,0.6,0}
    \definecolor{gray}{rgb}{0.5,0.5,0.5}
    \definecolor{mauve}{rgb}{0.58,0,0.82}
    \let\origthelstnumber\thelstnumber
    \makeatletter
    \newcommand*\Suppressnumber{%
      \lst@AddToHook{OnNewLine}{%
        \let\thelstnumber\relax%
         \advance\c@lstnumber-\@ne\relax%
        }%
    }

    \newcommand*\Reactivatenumber[1]{%
      \lst@AddToHook{OnNewLine}{%
       \let\thelstnumber\origthelstnumber%
       \advance\c@lstnumber\@ne\relax%
      }%
    }
    \makeatother

    \lstdefinelanguage{Kotlin}{
      comment=[l]{//},
      commentstyle={\color{gray}\ttfamily},
      emph={delegate, filter, first, firstOrNull, forEach, lazy, map, mapNotNull, println, return@},
      emphstyle={\color{OrangeRed}},
      identifierstyle=\color{black},
      keywords={abstract, actual, as, as?, break, by, class, companion, continue, data, do, dynamic, else, enum, expect, false, final, for, fun, get, if, import, in, interface, internal, is, null, object, override, package, private, public, return, set, super, suspend, this, throw, true, try, typealias, val, var, vararg, when, where, while},
      keywordstyle={\color{blue}\bfseries},
      morecomment=[s]{/*}{*/},
      morestring=[b]",
      morestring=[s]{"""*}{*"""},
      ndkeywords={@Deprecated, @JvmField, @JvmName, @JvmOverloads, @JvmStatic, @JvmSynthetic, Array, Byte, Double, Float, Int, Integer, Iterable, Long, Runnable, Short, String},
      ndkeywordstyle={\color{BurntOrange}\bfseries},
      sensitive=true,
      stringstyle={\color{black}\ttfamily},
    }

\makeatletter
\newcommand*\wrapletters[1]{\wr@pletters#1\@nil}
\def\wr@pletters#1#2\@nil{#1\allowbreak\if&#2&\else\wr@pletters#2\@nil\fi}
\makeatother

\newcommand{\todo}[1]{}%
\renewcommand{\todo}[1]{{\color{red} TODO: {#1}}}%

\newcommand{\tool}{\textsf{FlakeScanner}\xspace}
\newcommand{\dataset}{\textsf{FlakyAppRepo}\xspace}

\setcopyright{none}
\renewcommand\footnotetextcopyrightpermission[1]{}
\author{Zhen Dong}
\affiliation{National University of Singapore}
\email{zhen.dong@comp.nus.edu.sg}

\author{Abhishek Tiwari}
\affiliation{National University of Singapore}
\email{dcsabhi@nus.edu.sg}

\author{Xiao Liang Yu}
\affiliation{National University of Singapore}
\email{xiaoly@comp.nus.edu.sg}

\author{Abhik Roychoudhury}
\affiliation{National University of Singapore}
\email{abhik@comp.nus.edu.sg}

\begin{document}

%\title{\tool: Exposing Concurrency Flaky Tests in Android Apps}
\title{Concurrency-related Flaky Test Detection in Android Apps\footnotemark[1]}

\begin{abstract}
Validation of Android apps via testing is difficult owing to the presence of flaky tests. Due to non-deterministic execution environments, a sequence of events (a test) may lead to success or failure in unpredictable ways. In this work, we present an approach and tool \tool for detecting flaky tests through systematic exploration of event orders.  Our key observation is that for a test in a mobile app, there is a testing framework thread which creates the test events, a main User-Interface (UI) thread processing these events, and there may be several other background threads running asynchronously.
For any event $e$ whose execution involves potential non-determinism, we localize the earliest (latest) event after (before) which $e$ must happen. We then efficiently explore the schedules between the upper/lower bound events while grouping events within a single statement, to find whether the test outcome is flaky. We also create a suite of subject programs called \dataset to study flaky tests in Android apps. Our experiments on subject-suite \dataset demonstrate the efficacy of our flaky test detection. 
Our work is complementary to existing flaky test detection tools like Deflaker which check only failing tests. \tool can detect flaky tests among passing tests, as shown by our approach and experiments.
\end{abstract}

\maketitle
\footnotetext[1]{The extended version is available at \url{https://github.com/FlakyTest/FlakyTests} and will appear on ESCE/FSE2021.}

\section{Introduction}
\label{sec:intro}

GUI testing is notoriously difficult. A GUI test is brittle, a little change to underlying code could break the test. Even worse is a GUI test that passes sometimes and fails sometimes, without any change in the code, tests, or environment. These tests are called \emph{flaky tests}. Flaky tests occur often due to non-deterministic execution environments. Take an Android app, for example, that needs to connect to the internet. Based on the network stability, connection may take less or more time. If this kind of uncertainty is not properly dealt with, a test may pass when a connection is fast and fail for a slow connection, manifesting the flaky behavior.

Flaky tests frustrate developers and significantly hinder the testing automation. Developers often run tests to verify that their latest changes to a code repository did not break any previously working functionality, i.e., \emph{regression testing}. Ideally, a test failure would be due to the code changes and developers can focus on debugging these changes. Unfortunately, some test failures are not due to the code change but due to flakiness. Identifying whether a failure is due to code changes
or flaky tests may require tremendous efforts and slow down software development. Google statistics~\cite{google}
%\footnote{Flaky Tests at Google and How We Mitigate Them, May, 2016:\url{https://testing.googleblog.com/2016/05/flaky-tests-at-google-and-how-we.html}} 
show that in practice 84\% of transitions from pass to fail involve a flaky test in regression testing, which causes significant drag on software engineers. Facebook recently established the detection of flaky tests as a top research problem for software testing~\cite{mark}.

Today, the technology to detect flaky tests is not mature. Developers face a considerable amount of failures due to flaky tests in regression testing~\cite{understanding}. They struggle to distinguish them from the failures that are due to a recently introduced regression. A typical way to detect flaky tests is to rerun failing tests repeatedly (called \emph{RERUN}). Specifically, developers rerun each failing test multiple times after witnessing the failure. If some rerun passes, the test is marked flaky, otherwise it is marked as unknown. Despite being simple and often used in practice, RERUN is unreliable. Flaky tests are nondeterministic by definition, so there is no guarantee that the outcome of a flaky test will change within multiple reruns. Besides, RERUN is extremely costly: A GUI test typically executes much slower and may take minutes for a single run. %The overhead of RERUN may result in an unacceptable delay of integrating some relevant code changes (a code change is often disallowed to be integrated unless all tests pass).

DeFlaker \cite{deflaker} detects flaky tests with differential coverage analysis. A test is marked as flaky if the test fails and at the same time it did not cover any of latest code changes during regression testing. Unfortunately, DeFlaker only works on failing tests and cannot examine whether tests that passed are flaky. Flaky tests that pass in the current regression testing cannot be detected; thus, they remain in the test suite. %They can bring trouble in next regression testing.

%and sts that failed to pass. DeFlaker cannot diagnose whether a passing test is flaky and quarantine the test from test suite before regression testing occurs. Flaky tests that did not manifest flaky behavior can remain in the test suite without being discovered, and bring troubles later.

In this work, we propose a proactive approach to expose flaky tests of Android apps. Our approach takes a passing test as input and executes it in different possible execution environments. If there is an environment in which the test fails, the test is deemed flaky. With this approach, flaky tests can be exposed and quarantined from a test suite before regression testing occurs. Thus, failures due to test flakiness can be avoided in regression testing. This is with
the goal of reducing developer frustrations when they ``discover" that a test that they have been using all along is flaky! Instead, we check whether a test is flaky {\em before} introducing it into the test-suite.

\emph{ Observation.}  Android adopts the single GUI thread model in which events are processed by a UI thread sequentially. To avoid blocking the UI thread for responsiveness, a long-running task, such as accessing the internet, is offloaded to an async thread. Once the task is completed, the async thread updates the result by submitting an event (called an \emph{async event}) to the UI thread, which results in event racing. Under this model, event execution order varies from run to run due to non-deterministic execution environments. An app might behave normally for most orders but act in an unexpected way for certain orders.  This eventually leads to a phenomenon where a test passes sometimes and fails at other times.

\emph{Insight.} A flaky test can be exposed by exploring the space of event execution orders that may result from different execution environments. Instead of exhaustively running a test in all possible execution environments, we examine a test by exploring the space of feasible event execution orders to check whether there exists an event execution order in which the test fails. A test is deemed flaky if a failure is detected during exploration. Our insight is that Android apps often assign environment-related tasks such as accessing background service or the internet to async threads. App behaviors in different execution environments can be explored by scheduling events.

\emph{Challenges.} Computing the space of possible event execution orders is difficult. As explained above, event execution orders are determined when async threads submit the async events at runtime. A test run may involve many async threads which are launched from different layers including Android framework, third-party libraries, and the app under test itself. This poses a challenge to track them. Moreover, computing the possible event execution orders involves resolving event dependencies due to thread synchronization. A GUI test often runs in a separated thread maintained by a testing framework such as Espresso~\cite{espresso}. Thus, a thorough analysis of the interleavings between the app, the Android framework, and the testing framework is required. Existing techniques for Android apps focus mainly on finding specific concurrency bugs, e.g., CAFA~\cite{race} is limited to finding race errors due to use-after-free violations. Additionally, they only support app analysis and provide no support for testing framework.
%Existing techniques such as race detection CAFA and event scheduling RaceDroid only support app analysis and face challenges to perform such computation.

\emph{Solution.} We propose a system-level dynamic analysis to resolve thread synchronization dependencies. We run apps in the debug mode such that threads in the whole Android runtime can be monitored and controlled. Thread synchronization dependencies can be resolved by manipulating threads, e.g., suspending a thread to observe others.
%including  by taking control of the whole Android Runtime. We leverage the Android Debugger to control the entire ART environment. All threads originating from the Android framework, the testing framework, and the app under test can be monitored and manipulated as desired, e.g., suspending them.
%In such a way, we can observe the synchronizations among threads, timings of the submitted async event, and accurately compute the space of possible event execution orders.
%Besides, a GUI test often generates many events in the Android system, which leads to a huge space of possible event execution orders. Enumerating all of them is costly, considering a GUI test runs slowly and may take several minutes. To address this, we take a single statement in the test as a basic GUI operation and group events that are triggered by the same statement together. An async event is scheduled between event groups instead of all available events. Thus, the space of possible event execution orders can be significantly reduced.
Besides, a GUI test consists of multiple statements with various GUI operations such as a button click. These operations create multiple events to complete their execution, which leads to a huge space of possible event execution orders. Enumerating all of them is costly, considering a GUI test may run slowly and take several minutes. To address this, for each statement, such as a button click, we group the events it generates. Finally, we schedule the async events between these groups. %As a result, the space of possible event execution orders is significantly reduced.
%Our solution is embodied in the form of our flaky test detection tool \tool.

\emph{Experiments.} We evaluate our approach on \dataset which contains 28 widely-used Android apps including Firefox app from Mozilla. We detected 19 out of 24 previously known GUI related flaky tests. We analyzed their root causes and categorized them into three categories. Additionally, we discovered 245 flaky tests that were previously unknown. 

\emph{Contributions.} Our contributions can be summarized as follows.
\vspace{-1mm}
\begin{itemize}
	\item We propose a proactive approach  to expose flaky tests in a test suite. A flaky test can be exposed and quarantined before regression testing so that failures due to flaky tests can be avoided. It is the first approach that automatically detect flaky GUI tests for Android apps. %Our approach is embodied as a flaky test detection tool \tool.
	\item We develop a technique \tool which can control threads launched from different layers including apps, Android framework and testing framework, and perform a system level dynamic analysis to precisely resolve dependencies between events. %The tool \tool is publicly available.
	\item We collect a subject-suite that contains 28 widely-used apps with GUI tests that are from developers, called \dataset. To facilite future research on flaky tests, we make our tool \tool and subject-suite \dataset publicly available at:  {\it \url{https://github.com/FlakyTest/FlakyTests}}
\end{itemize}

\section{Android concurrency and testing} \label{sec:bk}

\begin{figure}[h]
	\center
	\includegraphics[width=0.7\linewidth]{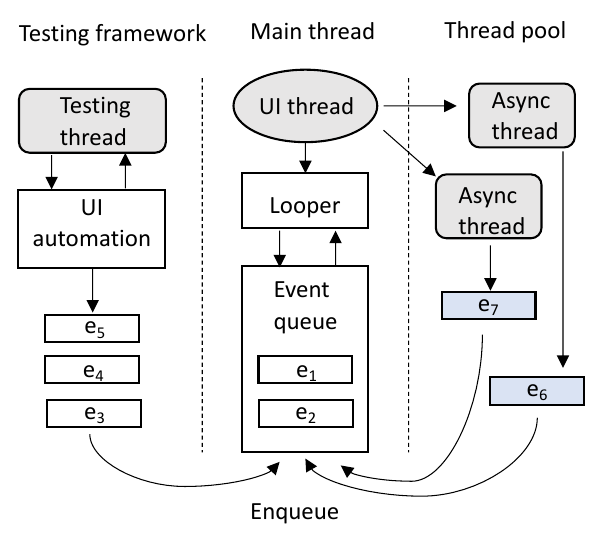}
	\caption{Android concurrency model \& testing framework.}
	\label{fig:concurrency}
\end{figure}

Figure~\ref{fig:concurrency} depicts the Android hybrid event-driven concurrency model as well as the Android GUI testing framework. Every app has a main thread (also called \emph{UI thread}), which maintains an event queue and a looper associated with the event queue. UI or system events that are generated by users or the system are added into the event queue. The looper dequeues events in a sequential order and dispatches them to corresponding handlers for processing. Android adopts the single-UI-thread model where only the main thread can access the GUI objects. To prevent non-responsive threads from blocking the GUI, long-running tasks such as network access are offloaded to async threads. Once these tasks are finished, async threads post an event marked in blue in Figure~\ref{fig:concurrency} to the main thread, which updates results to GUI objects. However, this concurrency model can lead to event racing. As shown in Figure~\ref{fig:concurrency}, the main thread and async threads run concurrently. When an async thread finishes a task and posts an async event is non-deterministic, depending on the current execution environment. Consequently, the order of events processed by the main thread is non-deterministic as well. In the example shown in Figure~\ref{fig:concurrency}, there are multiple possible orders which might occur in the execution such as <$e_1, e_2, e_6, e_3, e_4, e_5, e_7$> and  <$e_1, e_2, e_3, e_4, e_5, e_6, e_7$>.

Android provides testing frameworks for developers to write GUI tests, which simulate user interactions to exercise app functionalities. GUI tests run on physical devices or emulators and interact with UI interface to generate events.
%and are executed by a dedicated thread maintained by the testing framework (we call \emph{testing thread}).  The testing thread interacts with GUI interface to generate events.
%UIAutomation is an Android framework component that allows to inject user events such as clicking and tapping.
To achieve more reliable tests, testing frameworks  provide a set of mechanisms to synchronize test automation interactions with the user interface. For instance, when method \texttt{onView()} is invoked in a test, Espresso waits to perform the corresponding UI action or assertion until the event queue is empty and some async threads (e.g., AsyncTask instance) are terminated and user-defined resources are idling.
% since tests and apps are executed in different threads (see Figure~\ref{fig:concurrency})

\section{A Motivating Example}
\label{sec:exaple}
In this section, we use a simple example to explain how a flaky test occurs in Android apps and the challenges in detecting such flaky tests. The example comes from the \emph{RapidPro Surveyor} app for Android and related code snippets are shown in Listings~\ref{listing:CaptureLocationActivityTest}-~\ref{listing:googleapi}. As we see, the test (Listing~\ref{listing:CaptureLocationActivityTest}) first launches an activity (Listing~\ref{listing:CaptureLocationActivity}) that is used to capture location data  of the Android device. When created, the activity connects to the Google API client (line~\ref{lst:line2_2}, Listing~\ref{listing:CaptureLocationActivity}) and requests location data of the device. During the connection process, Google API client creates a few worker threads (refer Listing~\ref{listing:googleapi}) to complete the connection process. Finally, the Google API client accesses the location data and sends it to the activity. Then the test clicks a button on the activity to obtain the location data from the activity. In the end, the test checks the obtained data is not \texttt{NULL}.

Despite being simple, the test is a flaky test. As mentioned before, the test is executed in a testing thread, and the activity runs on the UIThread of the app,  and the operation that the Google API client obtains location data is executed on an async thread. Although the test uses \texttt{onView()} to synchronize GUI operations, the testing thread cannot synchronize with the async thread that fetches the location data. Thus, the async thread might update the location data to the activity before or after the testing thread checks the location data. If the checking occurs before the activity receives the location data, the test fails. Otherwise, the test passes. This leads to a phenomenon that the test passes for some times and fails for other times.

Detecting such flaky tests is difficult. First of all, a flaky test is "hiding" in the test suite and may pass in most execution environments. There is nothing different from other passing tests unless the test fails in the execution. The existing techniques of detecting flaky tests such as DeFlaker~\cite{deflaker}  apply on failing tests and cannot examine whether a passing test is flaky. Although  many existing techniques can detect concurrency bugs, they face challenges to detect a flaky test. As shown in the example, a test in Android apps is often run by a testing framework and many threads might come from the Android system. Detecting flaky tests requires to analyze not only app under test but also the testing frameworks. However, existing techniques typically focus on app analysis.   For instance, DroidRacer~\cite{droidracer} records execution traces and detects data race in apps by offline analyzing collected execution traces. ERVA~\cite{erva} takes a data race report which are generated by other tools like DroidRacer to verify whether the reported data race is true positive. AsyncDroid~\cite{asyncdroid} detects bugs in an app by exploring alternative execution orders of event handlers that are created by the app. None of them deals with analysis of testing frameworks and applies to flaky tests detection. This urgent need motivates us to develop a technique that can detect flaky tests for Android apps.

\begin{lstlisting}[ language=Java, belowskip=-0.8 \baselineskip, caption=CaptureLocationActivityTest (Espresso Thread), label={listing:CaptureLocationActivityTest}, float, escapeinside={*}{*}, basicstyle=\small]
*\label{lst:line1_11}*@Rule
*\label{lst:line1_11}*public ActivityTestRule<CaptureLocationActivity> rule = new ActivityTestRule<>(CaptureLocationActivity.class);
*\label{lst:line1_0}*@Test
*\label{lst:line1_2}*public void capture() {
*\label{lst:line1_3}*	onView(withId(R.id.button_capture))	*\Suppressnumber*		
							.check(matches(isDisplayed()))
							.perform(click());*\Reactivatenumber{}*
*\label{lst:line1_7}*	Instrumentation.ActivityResult result = rule.getActivityResult();
*\label{lst:line1_8}*   assertThat(result.getResultData(), is(not(nullValue())));
*\label{lst:line1_9}*	//...
*\label{lst:line1_10}*  }
}
\end{lstlisting}
			
\begin{lstlisting}[ language=Java, belowskip=-0.8 \baselineskip, caption=CaptureLocationActivity (Main Thread), label={listing:CaptureLocationActivity}, float, escapeinside={*}{*}, basicstyle=\small]
@Override
*\label{lst:line2_1}*protected void onCreate(Bundle bundle) {
	// ...
*\label{lst:line2_2}*	connectGoogleApi();
	//..
}
protected void connectGoogleApi() {
	googleApiClient = new GoogleApiClient.Builder(this)
									.addConnectionCallbacks(this)
									.addOnConnectionFailedListener(this)
									.addApi(LocationServices.API)
									.build();

	googleApiClient.connect();
}
\end{lstlisting}
\begin{lstlisting}[ language=Java, belowskip=-0.8 \baselineskip, caption=Zaau (Worker Thread), label={listing:googleapi}, float, escapeinside={*}{*}, basicstyle=\small]
//Worker thread asynchronous to the main thread
@WorkerThread
public void run() {
	zaak.zac(this.zagj).lock();
	try {
		if (!Thread.interrupted()) {
		this.zaan();
		return;
		}
	//...
}
\end{lstlisting}

\section{Overview}
\label{sec:approach}

\begin{figure}[t]
	\center
	\includegraphics[width=0.7\linewidth]{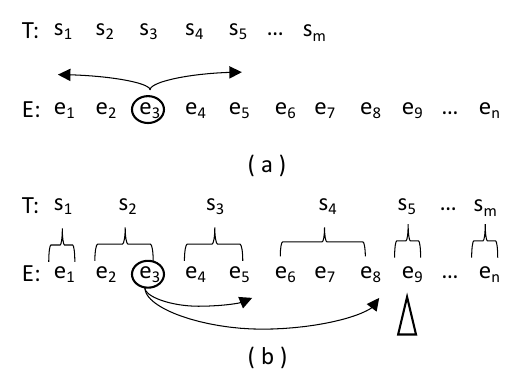}
		\caption{Computing event execution orders of a test.\label{fig:orders}
}
\end{figure}

Consider a GUI test $T$  consisting of a sequence of program statements <$s_1, s_2, s_3, s_4, s_5, ... s_m>$ in Figure~\ref{fig:orders} (a), its one possible execution trace is $E$ that consists of a sequence of executed events <$e_1, e_2, ...e_n$>, and $e_3$ is an async event (i.e., $e_3$ is generated by an async thread). For simplicity, we assume only one async event is generated for this test (in general, many async events are generated in a single test execution, which is considered in our approach). As previously stated, event execution order of the test might change for a different run due to non-deterministic execution environments. For instance, event $e_3$ might be executed before $e_2$ or after $e_7$, depending on how much time is needed for its corresponding thread to complete the task and post the event. Our goal is to compute how many possible event execution orders there are for this test and check whether there exists one order in which the test fails. If a test failure is detected, the test is deemed flaky.
% (and another order in which the test succeeds)

Apparently, the position of $e_3$ in the sequence of events is uncertain for a different test run. However, the space of possible positions of  $e_3$ should be constrained between certain two events due to dependencies resulted from thread synchronization. We assume the two events are $e_u$ and $e_v$. So $e_3$  cannot be executed earlier than  $e_u$ or later than  $e_v$,  no matter how the execution environment changes (e.g., network connection becomes slow). If $e_u$ and $e_v$ are localized, computing possible event execution orders for the test can be achieved.
%To localize $e_i$ and $e_j$ for the test, we need to perform a dynamic analysis to resolve thread synchronization dependencies.

\paragraph*{Computing space of event execution orders.} In event-driven programming, an event is designed for communication among multiple components and well-encapsulated. Event dependencies are typically handled over to other components like event handlers. Thus, it is challenging to identify such dependencies by capturing and analyzing events themselves. Our idea is to link events to statements in a test since all events are triggered by a test. We execute a test statement by statement and record all events triggered by each statement and build a map between them as shown in Figure~\ref{fig:orders} (b). As we see, $e_3$ is triggered by statement $s_2$. Consequently, $e_3$ cannot be executed earlier than the first event that is triggered by $s_2$. Thus, lower bound $e_u$ of the async event $e_3$ can be identified, i.e., $e_2$. Localizing upper bound $e_v$ of event $e_3$ involves identifying which events depend on $e_3$, i.e., events that occur only after $e_3$ is processed. Testing frameworks typically use thread synchronization to guarantee an event occurs before another. For instance in an Espresso test, a statement that invokes \texttt{onView()} method waits until specified threads or resources are idling; otherwise it refuses to be executed. Therefore,  event dependency analysis requires to resolve thread synchronization dependencies. As mentioned earlier, traditional program analysis faces challenges to perform such analysis because Android testing often involves many threads from third-party libraries, Android framework, and testing framework.  Existing analysis techniques hardly overcome those obstacles, as they are often restricted to analyzing the app code.

Addressing this challenge, we propose a \emph{what-if} analysis. Specifically, after the test is launched, we hook the async thread that posts $e_3$ at runtime, and suspend the async thread and let other threads free to go. At the same time, we monitor the testing thread and check at which statement of the test it stops and waits for the hooked thread to be completed. Suppose that the testing thread stops at statement $s_5$, we consider operations in $s_5$ depends on $e_3$ and these operations will not be executed until $e_3$ is processed. Thus, the first event $e_9$ triggered by $s_5$ is upper bound of $e_3$, that is, $e_3$ has to be executed before $e_9$. The idea behind our approach is what if it takes forever to compute $e_3$, operations in a test that depend on $e_3$ will not be triggered due to thread synchronization and those that do not depend on $e_3$ will be executed. So, events that depend on $e_3$ are identified dynamically.

\paragraph*{Reducing space of event execution orders.} Since the position of $e_3$ is between $e_2$ and $e_9$,  possible event execution orders for the test can be calculated by moving $e_3$ one position at a time, until reaching $e_9$, for instance, <$e_1, e_2, e_4, e_3, e_5, e_6,... e_n$>. In practice, the space of event execution orders can be huge because one GUI operation often triggers multiple events at runtime, {\em e.g.}, one click would generate "click down" and "click up" events. One test may trigger hundreds of events, which leads to a huge space of possible event execution orders. Exploring all of them is costly considering a GUI test runs slowly. Thus, we only consider event execution orders in which $e_3$ is located before the first event of a statement, i.e., between $e_5$ and $e_6$, and between $e_8$ and $e_9$ shown in Figure~\ref{fig:orders}(b). This is reasonable because app behavior is more likely to be influenced when an async event is executed after the execution of a statement is completed.

\paragraph*{Scheduling events.} %We schedule an event with threads manipulation.
Suppose, we are exploring an event execution order which in $e_3$ is executed prior to $e_9$ (see Figure~\ref{fig:orders} (b)). We first query the map between events and statement which is previously generated and identify which statement triggers $e_9$ (in this case it is statement $s_5$). Once the test run is launched, we hook the async thread that posts $e_3$ and suspend the thread such that the event $e_3$ cannot be posted. At the same time, we monitor the testing thread and check which statement is being executed by querying the program counter in the Android runtime. When the program counter reaches $s_5$, we suspend the testing thread and free the async thread that we suspended earlier.  After the async thread finishes the task, we free the testing thread to run. In this way, $e_3$ can be executed immediately before the first event $e_9$ of statement $s_5$.

\section{Methodology}
\label{sec:approach}

\begin{figure}[h]
	\centering
	\includegraphics[width=0.9\linewidth]{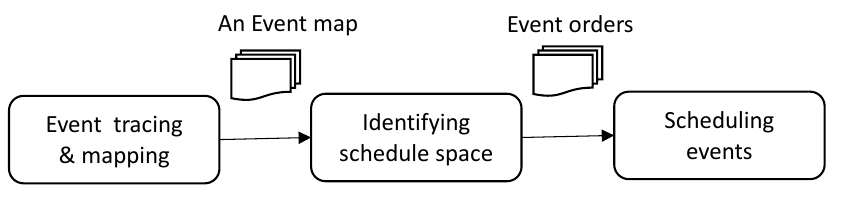}
	\caption{workflow of exposing flaky tests through systematic schedule exploration.}
	\label{fig:approach}
\end{figure}

Figure~\ref{fig:approach} shows the workflow of our approach. Given a test and app under test, it performs a concrete execution to trace events that the test generates and builds a map between  statements in the test and events that are triggered by these statements. Then the approach executes the test multiple times to compute possible schedules for async events and generates a set of event orders that might occur in execution environments. Finally, we explore these possible event execution orders for the test. During the exploration, if a test failure is detected, the test is identified as a flaky test.

\subsection{Event tracing and mapping}

Event tracing is often used in dynamic analysis of Android apps. It can be achieved by simply logging events that are generated at runtime. However, such techniques cannot fulfill our task. Event information ({\em e.g.}, event id) produced in logs is dynamically generated at Android runtime and changes in a different run. Our approach requires an event identifier which can be used to identify an event across different test runs. Async events that are identified during event tracing need to be hooked and scheduled in runs which are performed for event order exploration. This poses a challenge for existing techniques.

\paragraph*{Event identification.}  We identify an event based on interactions between the event and app under test at runtime.  Two events that are triggered in different test runs are considered as an identical program behavior if: (1) they are triggered by  the same test statement; (2) they are processed by a same sequence of methods at runtime. For instance, a \texttt{pressDown} event is associated with an identifier which is constructed with line number of the statement that triggers the event and signatures of a sequence of methods that process the event.  This practice of event identification comes from our investigation of the Android framework. Events are widely used for thread communication and managed by \texttt{Handlers} associated with threads. Events are dealt via different \texttt{Handlers} according to where events come from. % i.e., different events are handled with different methods.

\begin{algorithm}[t]
  \small

	\caption{Event tracing and mapping.\label{alg:mapping}}

  \SetKwFunction{run}{runTest}
  \SetKwFunction{launch}{launchApp}
  \SetKwFunction{monitor}{attachHandler}
  \SetKwFunction{runtest}{launchTest}
  \SetKwFunction{getLineNum}{getLineNum}
  \SetKwFunction{getEventQ}{getEventQ}
  \SetKwFunction{runStatement}{runStatement}
  \SetKwFunction{isDone}{isAllEventHooked}
  \SetKwFunction{getTriggeredEvent}{getEvent}
  \SetKwProg{proc}{Procedure}{}{}

%  \SetAlgoVlined
%  \SetAlgoLined
	\proc{\run{App A, Test T, Android ART}} {
	 \launch($A$, $ART$)\;
	 $ARTHandler \leftarrow $ \monitor($A$, $ART$)\;
	 $List \leftarrow \emptyset$ \tcp*{ storing pairs of a statement and events}
	 \runtest($A$, $T$, $ART$)\;
         \tcp{UI thread's event queue}
	 $eventQ \leftarrow \getEventQ(ARTHandler) $\;

	 \For { \textbf{s} in $T$} {
		 $\runStatement(ARTHandler, s) $\;
		 $n \leftarrow \getLineNum(ARTHandler, s) $\;
		\While{True} {
	              $E\langle isAsync,sq\_m\rangle \leftarrow \getTriggeredEvent(ARTHandler) $\;
	      	      \eIf {E != Null} {
		
		           $List \leftarrow List \cup \{\langle isAsync, sq\_m, n\rangle\} $	
	              }
	             {
		          \If {isEmpty($eventQ$)}{
				  \textbf{break} \;
		       }
		   }
	 }

   }
   \Return $List$
}
\end{algorithm}

\paragraph*{Tracing and mapping.} Algorithm~\ref{alg:mapping} outlines the procedure of event tracing and mapping. It first launches app under test and takes control of Android runtime in which the app runs with a module called \emph{ARTHandler}. ARTHandler runs the input test $T$ in the testing thread and executes statements one by one. When one statement is executed, ARTHandlers monitors the event queue of the UI thread and hooks injected events. For each event, ARTHandler records the tuple $\langle isAsync, sq_m\rangle$  where $isAsync$ denotes  whether it is an async event and $sq_m$ denotes the signatures of a sequence of methods that have processed the event. This tuple, along with the line number of the statement that is being executed forms the identifier of the event, which is stored in a list. As stated before, a statement in the test might launch long-running tasks which are executed in async threads. Async events might take long time to be posted. To not miss async events that are triggered by a single statement, we keep hooking events until two criteria are satisfied: (a) there are no new events and (b) the event queue of the UIThread is empty, which often indicates the system is not running tasks. This practice is also used in the Espresso testing framework. A map between statements and events is stored in $List$ and returned.

\subsection{Identifying event schedule space}

\begin{algorithm}[t]
  \small

  \caption{Event scheduling exploration.\label{alg:exploring}}

  \SetKwFunction{run}{explore}
  \SetKwFunction{launch}{launchApp}
  \SetKwFunction{monitor}{attachHandler}
  \SetKwFunction{runtest}{launchTest}
  \SetKwFunction{getLineNum}{getLineNum}
  \SetKwFunction{getEventQ}{getEventQ}
  \SetKwFunction{getTestingTh}{getTestingThread}
  \SetKwFunction{suspend}{suspend}
  \SetKwFunction{hookAsyncTh}{hookAsyncThread}
  \SetKwFunction{runStatement}{runStatement}
  \SetKwFunction{isWaiting}{isWaiting}
  \SetKwFunction{isDone}{isAllEventHooked}
  \SetKwFunction{getTriggeredEvent}{getEvent}
  \SetKwFunction{isAsync}{isAsync}	
  \SetKwFunction{update}{updateEvent}
  \SetKwFunction{setUpperBound}{setUpperBound}
  \SetKwProg{proc}{Procedure}{}{}
%  \SetAlgoVlined
%  \SetAlgoLined
	\proc{\run{App A, Test T, Android ART, EventMap List}} {
	\For{\textbf{e} in $List$}{
		\If{\isAsync(e)}{
		 \launch($A$, $ART$)\;
	 	 $ARTHandler \leftarrow $ \monitor($A$, $ART$)\;
	         \runtest($A$, $T$, $ARTHandler$)\;
		 $TH_{test} \leftarrow $ \getTestingTh($ARTHandler$)\;
		 $TH_{async} \leftarrow $  \hookAsyncTh($ARTHandler$)\;
		 \suspend($ARTHandler,TH_{async}$) \;
		 \While{True}{
			 \If {\isWaiting($TH_{test}$)}{
				 \textbf{break} \;
		          }
		 }
		 $m \leftarrow $ \getLineNum($ARTHandler,Th_{test}$)\;
	     }
	     %$e \leftarrow $ $<e, m>$ \;
	     \tcp{$m$ is set as upper bound along with $e$}
	     \setUpperBound($e$, $m$)  \;
	     \tcp{update event $e$ (with upper bound $m$) in List} 
	     \update($List$, $e$) \;
	     \textbf{break} \;
	 }

   \Return $List$
}
\end{algorithm}

To compute possible event execution orders, we perform a what-if dynamic analysis to resolve event dependencies that are caused by thread synchronization in apps and testing frameworks. Algorithm~\ref{alg:exploring} shows the procedure of resolving event dependencies. It takes the event trace $List$  generated in the previous step as input. For each async event $e_i$ in $List$, the algorithm launches the test and starts to hook event $e_i$. Once hooked, the algorithm suspends thread that posts $e_i$ such that $e_i$ can be posted. Meanwhile, it keeps checking status of the testing thread. If the status of the testing thread is \texttt{WAITING}, it considers the testing thread is performing thread synchronization with threads in the app and waiting for $e_i$ to be executed. Thus, we consider the statement $s$ that is being executed in the testing thread  attempts to trigger an event (saying $e_j$)  which depends on $e_i$.
%e.g.a statement invoking method \texttt{onView()}.
Therefore,  the schedule space of $e_i$ is bounded by $e_j$, i.e., the first event that is triggered by $s$. So statement $s$ is identified as the upper bound of schedule space of async event $e_i$. Statement $s$ is recorded and set as the upper bound of event $e_i$. When the upper bound is set, $e_i$ is restored to $List$.  In the end, schedule spaces for all async events in $List$ are identified and recorded.

\subsection{Scheduling events}
Schedule space of each async event in the event trace $List$ is identified in previous steps. Now we explore event orders during test execution. An async event $e_i$ can be simply represented by a triple <$id, n, m$> where $n$ and $m$ are bounds of the schedule space of $e_i$. Specifically,  $n$ is the index of the statement in the test that triggers $e_i$,   and $m$ is the index of the statement that triggers the upper bound event of $e_i$ .

Similar to schedule space identification, we can schedule $e_i$ by operating threads. We first hook event $e_i$  after the test is launched and suspend the thread that posts $e_i$. Then, we free the testing thread and monitor whether the statement that is being executed is statement $m$. Once statement $m$ is reached, we suspend the testing thread and free the suspended thread to post $e_i$. After the async thread is terminated or idling, i.e., event $e_i$ has been posted, we free the testing thread. In such a way, event $e_i$ can be executed prior to statement $m$. In next test run, we schedule $e_i$  to be executed prior to statement $m-1$, until all statements between $n$ and $m$ are explored. This procedure is repeated for each async event in $List$ so that the space of possible event execution orders can be systematically explored for the test.

\subsection{Optimization}

In Algorithm~\ref{alg:exploring}, each async event requires one test run for schedule space identification. If $n$ async events are generated for a test,  we need to run the test $n$ times to identify the space of event execution orders, which is costly. To address this issue, we perform an optimization on schedule space identification. When the schedule space of an async event is identified, instead of terminating the execution, we continue executing the test to identify schedule spaces for subsequent async events so as to reduce the number of test runs. In this case, we may suspend multiple async threads at the same time and release more than one async threads to resolve a thread synchronization dependency, which results in a group of async events having the same upper bound (latest time). We only rerun the test for this case to identify schedule space of each of them. Thus, the total number of test runs can be significantly reduced.

\section{Implementation}
\label{sec:impl}

%  System architecture
%  \begin{itemize}
% 	 \item Java debugger
% 	 \item Threads controlling
% 	 \item event enqueque hooking
% \end{itemize}

%\begin{figure}[h]
	% \center
%	\includegraphics[width=\linewidth]{figures/System-Arch.pdf}
%	\caption{System Architecture}
%	\label{fig:system}
%\end{figure}

Our system is implemented in Scala and runs on a computer that connects a physical Android device or an emulator. Unlike existing techniques, it requires no instrumentation on apps or the Android framework, nor any modification to the Android framework, and can be easily adapted to different versions of Android.

\paragraph{Taking control of Android Runtime.} We leverage the Android debug mode to control the Android runtime. The Android framework allows to run an app in the debug mode. In this mode, we interact with the Android runtime using ADB to remotely monitor the app state and manipulate the thread executions, e.g., performing the execution step by step.

\paragraph{Hooking events.} Android adopts the event-driven model, in which each app has an event queue for storing events that occurred and processes them one by one. In the debug mode, we are allowed to set a breakpoint at method \texttt{enqueueMessage()} which is in charge of enqueuing events. Whenever the method is invoked, our system is informed and performs predefined operations such as suspending the event-posting thread. In such a way, our system can hook any event that occurs in the Android runtime before the event is posted into the event queue.

\paragraph{Operating threads.} In the debug mode, we are allowed to inspect threads that are running in the Android runtime and check their statues and operate them by sending commands, e.g., sending a command to release a suspended thread. So the UIthread and testing thread can be identified during a test execution.  A breakpoint is inserted at each test statement such that we can fully monitor and control the testing thread including querying the index of statement that is being executed and executing step by step. We also can examine the stack frames of a thread to check executed methods in the thread. Such data is used to identify an event.

%We have faithly implemented our approach in the Scala programming language. The system contains 2 main running components as shown in figure \ref{fig:system}.
%Our system controls the Android system framework and inspects its state through Android Debug Bridge (adb) and the interfacing is done via ddmlib.
%After the communication channel between our system and the Android system is established, our system creates JDWP(Java Debug Wire Protocol) packets to manipulate the ART runtime instance corresponding to the app under test. The manipulation includes event hooking and thread suspension. We intercept all events sent to the main UiThread looper by setting a backpoint at the enqueueMessage method of the main event queue instance. By default, the threads enqueuing new events are suspended until they are determined to be unsuspended by our described approach. Upon unsuspension, the message will be enqueued and therefore can be processed.
%The detection of whether the event queue is empty is achieved by setting a breakpoint at the IdleListener processing code block. Thus, whenever the message queue is idle, our system can be informed to process promptly and accordingly.

\section{Evaluation}
\label{sec:evaluation}
We perform evaluation on the effectiveness of \tool in detecting flaky tests that reside in test suites of real world Android apps. Our evaluation aims to address the following research questions:

\begin{itemize}
	\item Can \tool  examine and detect known \emph{flaky tests}? % and detect them as flaky?
	\item How does \tool compare with existing techniques in terms of number of detected flaky tests?
	\item Can \tool  be used to discover new flaky tests in apps?
	%\item How is the performance of our technique in terms of times running a test for such detection compared to RERUN?
\end{itemize}
\subsection{Subject apps}
Android app testing has been heavily explored. There are various benchmarks used in evaluating the effectiveness of automated testing of Android apps such as AndroTest~\cite{androtest} which contains 68 open source Android projects and the benchmark~\cite{taoxie} from Wang et al. which contains 68 industrial Android apps. However, few apps from those benchmarks come with a test suite from developers, let alone GUI tests.  On the other hand,  test flakiness is an urgent challenge, especially for GUI tests. Unfortunately, there are no Android app benchmarks to support test flakiness research.

Given the pressing need, we developed the first subject-suite \dataset which is used to study GUI test flakiness. It contains 28 widely-used Android apps including Mozilla Firefox Lite and WordPress as shown in Table~\ref{table:subapps}. There are more than 5000  Android instrumentation tests from developers that run on physical devices and emulators. %Out of them there are more than 2000 GUI tests and 156 of them are annotated with flaky tests by developers. 

The  challenge we face in building this data set is that publicly available Android projects rarely have tests from developers. To overcome this challenge,  we collect Android projects with the following strategies. First, we search well-known Android projects like Firefox in Github and select any project in which there exist tests under folder “../src/androidTest”. Second, we search the label “@FlakyTest” in Github and on the Google website, and select any Android project in which at least one of instrumented tests is labelled “@FlakyTest”. Tests labelled “@FlakyTest” in a test suite are flaky tests reported by developers.

%To facilitate future research on flakiness of GUI tests, we made the benchmark publicly available.

%{\em Upon publication of our paper, we will publish the tool \tool open source, from the same website where
%we have already made \dataset available:} \url{https://github.com/FlakeShovel/FlakeShovel}
%{\em Our tool \tool and subject-suite \dataset are availabe at the anonymous link} \url{https://github.com/FlakeShovel/FlakeShovel}. 

\subsection{ Experiment setup}
\label{sec:setup}

We conduct two studies to answer above research questions. Study 1 addresses RQ1 and RQ2 and study 2 addresses RQ3. For study 1, we evaluate \tool on GUI tests in \dataset (that have been annotated as flaky tests by developers) to check whether \tool can detect such flaky tests. First, we select all tests in the benchmark that are marked as flaky and exclude tests that are not GUI tests ({\em e.g.}, tests for database operations), then execute them on Android emulators. The passing tests are used in study 1. In the end, 24 GUI tests are collected from 6 apps, which is shown in Table~\ref{tab:rq1}. \tool and RERUN execute each of them. If a test failure occurs during execution, the flaky test is considered to be successfully detected. The results of each test is recorded for analysis. For study 2, we exclude tests that are used in study 1 and execute the remaining Android instrumentation tests on emulators and the passing tests are selected for evaluation. Eventually, 1444 tests are obtained from the 28 apps and are executed by \tool. If a test failure is detected, \tool reports the test as a flaky test.

We conducted experiments on a physical machine  with 64 GB RAM and a 56 cores Intel(R) Xeon(R) E5-2660 v4 CPU, running a 64-bit Ubuntu 16.04 operating system. Each execution instance runs in a Docker container to minimize the potential inference between running instances. App under test runs on an Android 9 (x86) emulator. One execution instance is for one test case for which the Android emulator is initialized to a fresh state at the begining to provide a clean testing environment.

\begin{table}[t]
\caption{Subject Apps}
\centering
\begin{adjustbox}{width=\linewidth}
	\small
 \begin{tabular}{||l | l | l | l |l ||}
 \hline
 App Name & Version & \#LOC & \#Stars  & Category\\ [0.5ex]
 \hline
 \hline
 Amaze File Manager & 3.4.3 & 92.2k & 2.8k & Tools \\
 \hline
 Youtube Extractor & 2.0.0 & 2.7k & 519 & Video Players \\
 \hline
 AntennaPod & 1.8.1 & 102.6k & 2.7k & Music \& Audio \\
 \hline
 CameraView & 2.6.1 & 40.5k & 2.9k & Photography \\
 \hline
 Catroid & 0.9.69 & 457.5k & 0 & Education \\
 \hline
 City Localizer & 1.1 & 4k & 0 & Travel \& Local \\
 \hline
 ConnectBot & 1.9.6 & 119.7k & 1.4k & Communication \\
 \hline
 DuckDuckGo & 5.43.0 & 211.3k & 1.2k & Tools \\
 \hline
 Espresso & 1.0 & 17.3k & 1.1k & Maps \& Navigation \\
 \hline
 Firefox Lite & 2.18 & 1598.4k & 212 & Communication \\
 %\hline
% FlexBox & 2.0.1 & 29.2k & 15.2k & Libraries \& Demo \\
 \hline
 GnuCash & 2.4.0 & 90.1k & 1k & GnuCash \\
 \hline
 IBS FoodAnalyzer & 1.2 & 26.1k & 0 & Health \& Fitness \\
 \hline
 Google I/O & 7.0.14 & 73.5k & 19.6k & Books \& Reference \\
 \hline
 Just Weather & 1.0.0 & 5.9k & 65 & Weather \\
 \hline
 KeePassDroid & 2.5.3 & 159.7k & 1.2k & Tools \\
 \hline
 KickMaterial & 1.0 & 79.1k & 1.6k & Crowdfunding \\
 \hline
 KISS Launcher & 3.11.9 & 27.2k & 1.4k & Personalisation \\
 \hline
 MedLog & 1.0 & 65k & 0 & Medical \\
 \hline
 Minimal To Do & 1.2 & 27.5k & 1.8k & Productivity \\
 \hline
 MoneyManagerEx & 02.14.994 & 170k & 265 & Finance \\
 \hline
 My Expenses & 3.0.7.1 & 1835.6k & 248 & Finance \\
 \hline
 NYBus & 1.0 & 6.9k & 272 & Transport \\
 \hline
 Omni Notes & 6.0.5 & 105.9k & 2k & Productivity \\
 \hline
 OpenTasks & 1.2.4 & 448k & 724 & Productivity \\
 \hline
 ownCloud & 2.14.2 & 333.7k & 2.9k & Productivity \\
 \hline
 Sunflower & 0.1.6 & 5.3k & 10.1k & Gardening \\
 \hline
 Surveyor & 13.3.0 & 290.4k & 13 & Communication \\
 \hline
 WordPress & 14.2-rc-2 & 461.7k & 2.3k & Productivity \\
 \hline
\end{tabular}
\end{adjustbox}
\label{table:subapps}
\end{table}

\subsection{RQ1: Efficacy}
\label{sec:rq1}

\begin{table*}[t]
	\caption{Results on known flaky tests by \tool and RERUN.}
	\label{tab:rq1}
	
	\begin{adjustbox}{width=0.95\linewidth}

\begin{tabular}{|l|l|l|cccccr|ccr|}
	\hline
\multirow{2}{*}{Test Id} & \multicolumn{1}{c|}{\multirow{2}{*}{App:Framework}}                                           & \multicolumn{1}{c|}{\multirow{2}{*}{Method name}}              & \multicolumn{6}{c|}{\tool}                 & \multicolumn{3}{c|}{RERUN} \\
                         & \multicolumn{1}{c}{}                                                                         & \multicolumn{1}{|c|}{}                                          & \#Op & \#Events & \#Run & Time(s) & Ctg & Succ & \#Run  & Time(s)  & Succ  \\ \hline
1                        & \multirow{4}{*}{Surveyor: Espresso}                                                          & capture                                                       & 1    & 58       & 2     & 48      & C3  & \checkmark    & 2      & 5        & \checkmark     \\
2                        &                                                                                              & twoQuestions                                                  & 15   & 104      & 2     & 48      & C1  & \checkmark    & 20     & 264      &       \\
3                        &                                                                                              & multimedia                                                    & 9    & 233      & 2     & 158     & C2  & \checkmark    & 20     & 113      &       \\
4                        &                                                                                              & contactDetails                                                & 4    & 104      & 2     & 48      & C1  & \checkmark    & 20     & 276      &       \\ \hline
5                        & \multirow{4}{*}{\begin{tabular}[c]{@{}l@{}}Android Youtube \\ Extractor: Junit\end{tabular}} & testEncipheredVideo                                           & 3    & 4        & 2     & 24      & C3  & \checkmark    & 2      & 9        & \checkmark     \\
6                        &                                                                                              & testUnembeddable                                              & 5    & 7        & 3     & 22      & C1  &      & 20     & 23       &       \\
7                        &                                                                                              & testAgeRestrictVideo                                          & 5    & 5        & 3     & 12      & C1  &      & 20     & 21       &       \\
8                        &                                                                                              & testUsualVideo                                                & 3    & 5        & 3     & 14      & C1  &      & 20     & 19       &       \\ \hline
9                        & \multirow{5}{*}{\begin{tabular}[c]{@{}l@{}}MyExpenses: \\ Espresso\end{tabular}}             & testScenarioForBug5b11072e6007d59fcd92c40b                    & 4    & 861      & 2     & 100     & C1  & \checkmark    & 20     & 153      &       \\
10                       &                                                                                              & editCommandKeepsListSize                                      & 2    & 101      & 2     & 54      & C1  & \checkmark    & 20     & 168      &       \\
11                       &                                                                                              & cloneCommandIncreasesListSize                                 & 2    & -        & -     & -       & -   &      &    20    & 156         &       \\
12                       &                                                                                              & changeOfFractionDigitsWithUpdateShouldKeepTransactionSum      & 5    & 991      & 2     & 170     & C2  & \checkmark    & 20     & 142      &       \\
13                       &                                                                                              & changeOfFractionDigitsWithoutUpdateShouldChangeTransactionSum & 5    & 991      & 2     & 174     & C2  & \checkmark    & 20     & 135      &       \\ \hline
14                       & \multirow{4}{*}{\begin{tabular}[c]{@{}l@{}}Firefox Lite: \\ Espresso\end{tabular}}           & saveImageThenDelete\_imageSaveAndDeleteSuccessfully           & 3    & 627      & 2     & 164     & C2  & \checkmark    & 20     & 230      &       \\
15                       &                                                                                              & dismissMenu                                                   & 3    & 165      & 2     & 76      & C1  & \checkmark    & 20     & 171      &       \\
16                       &                                                                                              & turnOnTurboModeDuringOnBoarding\_turboModeIsOnInMenu          & 8    & 89       & 2     & 40      & C1  & \checkmark    & 1      & 7        & \checkmark     \\
17                       &                                                                                              & changeDisplayLang                                             & 5    & 189      & 2     & 72      & C1  & \checkmark    & 3      & 33       & \checkmark     \\ \hline
18                       & \multirow{6}{*}{\begin{tabular}[c]{@{}l@{}}AntennaPod: \\ Robotium\end{tabular}}             & testGoToPreferences                                           & 1    & 107      & 2     & 56      & C1  & \checkmark    & 4      & 22       & \checkmark     \\
19                       &                                                                                              & testClickNavDrawer                                            & 7    & 132      & 2     & 60      & C1  & \checkmark    & 1      & 5        & \checkmark     \\
20                       &                                                                                              & PlaybackSonicTest\#testContinuousPlaybackOnMultipleEpisodes   & 4    & 177      & 2     & 142     & C2  & \checkmark    & 20     & 254      &       \\
21                       &                                                                                              & PlaybackSonicTest\#testContinousPlaybackOffMultipleEpisodes   & 3    & 176      & 2     & 140     & C2  & \checkmark    & 20     & 279      &       \\
22                       &                                                                                              & PlaybackTest\#testContinousPlaybackOffMultipleEpisodes        & 3    & 175      & 2     & 142     & C2  & \checkmark    & 20     & 276      &       \\
23                       &                                                                                              & PlaybackTest\#testContinuousPlaybackOnMultipleEpisodes        & 4    & 162      & 2     & 140     & C2  & \checkmark    & 20     & 241      &       \\ \hline
24                       & \begin{tabular}[c]{@{}l@{}}CameraView: \\ Espresso\end{tabular}                              & preview\_isShowing                                            & 1    & -        & -     & -       & -   & -    &  20      & 86         &       \\ \hline

sum/ave                  &                                                                                              &                                                               & 4.4  & 248      & 2     & 83      &     & 19   & 15     & 127       & 6   \\ \hline 
\end{tabular}
	\end{adjustbox}
\end{table*}

Table~\ref{tab:rq1} shows results of \tool on the data set of known flaky tests. The first column indicates test Ids, the second column shows app names and testing frameworks used in apps, and the third column indicates test method names. Column "\#Op" represents how many statements in a test which perform thread synchronization during testing. This is computed by manually counting synchronization operations such as \texttt{waitFor()}, \texttt{await()} as well as \texttt{onView()}, \texttt{onData()} in the Espresso framework. Column "\#Events" indicates the number of events observed by \tool during detection. Column "\#Run" shows times the number of times a test is executed for flakiness detection. Column "Time" reports the time that is used to detect a flaky test. Column "Ctg" indicates which category the flaky test belongs to in root cause analysis (we identify four categories C1-C3, as mentioned in the following). Column "Succ" indicates whether the test is identified as a flaky test by \tool.

As we see in Table~\ref{tab:rq1}, \tool successfully detected 19 flaky tests. For test 6, 7, and 8, \tool could execute them but failed to identify them as flaky tests. Code inspection shows these tests are flaky due to using a unsophisticated synchronization mechanism, i.e., waiting for a fixed amount of time for asynchronization operations. These 3 tests extract meta data for given videos on the internet and specify 10 seconds waiting for accessing the internet. If it takes more than 10 seconds to connect the internet, they will fail.  \tool can monitor thread synchronization between testing frameworks and apps and stops delaying async events once this synchronization occurs. Thus, these tests passed without being identified as flaky tests. For test 11 and 24, \tool failed to execute them due to configuration issues, e.g., \emph{preview\_isShowing} from \texttt{CameraView} app passes in API 21 and fails for any of the later APIs: This test usage a view called \emph{TextureView}, which was updated in later Android APIs.

Testing frameworks often provide  various mechanisms to avoid test flakiness, e.g., Espresso uses method \texttt{onView()} to synchronizes view operations with the UIThread. Tests in Table~\ref{tab:rq1} are developed with such test frameworks. Why are they still flaky? To answer this question, we perform an empirical study on root causes of these flaky tests. The root causes are classified into 3 categories:

\textbf{Category 1} (C1): Tests are flaky due to non-deterministic execution environments. Apps often interact with background services or resources and exchanges data. For some reason (e.g., being used in other computation), these services or resources may be unable to respond in time, which leads to a \texttt{Timeout} exception in the UI thread or testing frameworks and causes a test failure eventually. As shown in Table~\ref{tab:rq1}, it is the most common root causes and 12 tests belong to this category. Unfortunately, testing frameworks cannot handle such asynchronism that occurs in the execution environment though they provide mechanisms to synchronize GUI operations or pre-defined resources.

\textbf{Category 2} (C2): A test expects an implicit event execution order which may not always occur in the execution. Events are not only used for data exchange between threads but also used to perform operations, e.g., an intent is often used to launch an activity. An event execution order change resulted from async threads can lead to a different app behavior such as the soft keyboard disappearing late, which leads to a test failure. In our study, 8 cases belong to this category.

\textbf{Category 3} (C3): Flaky tests are caused by data race between the testing thread and threads in apps. In many cases, data that is used to check app behavior by a test is produced asynchronously, i.e., by a background thread. The data can be updated late for sometimes and the test checks "old" data, which lead to a test failure. We have 4 such as cases in our study.

In summary,  despite testing frameworks’ support to eliminate flakiness, flaky tests still occur due to non-determinism from execution environments (C1 and C2) and developers omitting certain cases resulted from thread concurrency (C3).

%though testing frameworks provide  mechanisms to eliminate flaky tests.

\subsection{RQ2: Comparison with existing techniques}
\label{sec:rq2}

\paragraph{Comparison with RERUN} RERUN is widely used to examine whether a test is flaky. A failed test is deemed flaky if the test passes in multiple reruns. The approach works in our setting as well. A passing test is deemed flaky if the test fails in multiple reruns. We take RERUN as the base line tool and compare it with \tool. In our experiment, we rerun a test for 20 times. If a failure is detected, the test is identified as a flaky test and the execution is terminated.

As shown in Table~\ref{tab:rq1}, for most cases, \tool successfully detects a flaky test during the second run. Totally 19 flaky tests are detected and all of them are detected less than 3 minutes. RERUN successfully detects 6 flaky tests. Two of them are detected in the first run. The others are detected in 4 runs. In terms of execution time of a single test, RERUN runs faster than \tool since \tool takes time for dynamic analysis. However, \tool detects much more flaky tests than RERUN.

With regards to the timing comparison with RERUN, note that RERUN was run only 20 times per test. As a result, when we report that RERUN took 127 seconds on average, it is a gross underestimation of the actual time taken by RERUN.
%For example, RERUN took 496 runs to detect test .   

Overall \tool detects most flaky tests in the second run i.e., identifying event schedule space phase. The experiments also show that our optimization on schedule space identification is effective.  Schedule space identification involves maximally delaying an async event,  which most likely triggers a test-flaky failure. We compute schedule spaces of multiple async events at the same time. A flaky test most likely fails in this phase.  Therefore, this strategy can significantly reduce the number of runs to detect a flaky test.

\paragraph{Comparison with race-detection techniques.} Event race detection techniques DROIDRACER~\cite{droidracer}, ERVA~\cite{erva}, EVENTRACER~\cite{scalable}, and CAFA~\cite{race} run on a modified Android framework 4.3 or 4.4. We face  incompatibility issues to evaluate those techniques on the collected apps because many apps target Android frameworks with a higher version (we use Android framework 9.0 in the evaluation). Therefore, we perform a qualitative comparison analysis between \tool and race-detection techniques.

{\em False positives.} Flaky test detection leveraging data race detection techniques will depend on an accurate computation of the happens-before relation. However, data race detection techniques capture happen-before relations by monitoring event operations in UIThread. Event dependencies due to synchronization between testing framework (like Espresso) and  UIthread (app under test) will not be captured. This leads to an underestimation of the happens-before relation.  As an example event e1 can denote a button appearing in the screen due to an async thread completing computation, event e2 can be a button click in testing framework, and e1 happens-before e2. Such happens-before edges are dropped by race detectors. Flaky test detection leveraging data race detectors may lead to many false positives among the flaky tests reported. In contrast, \tool can precisely detect synchronization between testing framework and UIthread to avoid such false positives.

{\em Maintainability.} Race-detection techniques often require a modified Android framework to capture happen-before relations and can struggle from fast evolution of Android frameworks. By contrast, \tool can be used in different Android framework versions since \tool  requires no modification to Android framworks and the debug mode that \tool relies on is supported by most Android frameworks.

\subsection{RQ3: Real-world flaky detection}
\begin{figure}[t]
	%\center
	\includegraphics[width=0.85\linewidth,left]{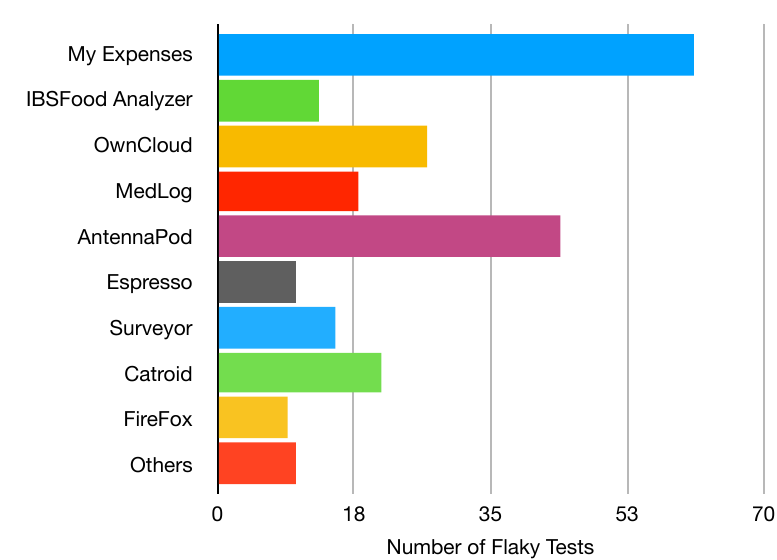}
		\caption{Flaky test distribution.\label{fig:distribution}
}
\end{figure}

To validate the effectiveness of \tool on discovering new flaky tests, we ran \tool on 1444 tests in \dataset which are not marked as flaky. The results show \tool is effective in discovering new flaky tests. Out of 1444 tests, \tool successfully detected 245 flaky tests. Figure~\ref{fig:distribution} shows distribution of the detected flaky tests among apps. \tool discovered the most flaky tests in app My Expenses (61 flaky tests) and  less flaky tests in FireFox. 
%This is not surprising since Firefox is actively maintained and My Expenses heavily uses Android instrumentation tests during development and these tests may not be well maintained. 
We also collected statistics on error messages of the failures of these flaky tests, which is shown in Table~\ref{table:exception}. The most common error message is "Waited for the root of the view hierarchy to have window focus" and the next is "No views in hierarchy found matching". In other words, most failures are related to mismatch between GUI operations and app state.

\subsubsection{Manually Checking Ground Truth}

To further validate the detected results, we manually investigated 20 randomly selected cases among 245 reported tests.  This is to manually check whether the tests are actually flaky.

The investigation shows these tests usually pass, however, they fail only for one or some corner cases. We leveraged the event orderings discovered by \tool to identify these corner cases. Two of the authors then studied the effect of those event orders and verified that these tests were indeed flaky. As it turns out during this manual process the two authors were in agreemnt and there was no disagreement which needed to be resolved. We then reported these cases to the corresponding developers with the detailed reports on how to reproduce them.  

At the time of writing of the paper, we got 11 out of the 20 test cases confirmed as flaky tests. Five tests are still under investigation by developers, and we did not hear back from developers for four tests.

%We leveraged the event orderings discovered by \tool to identify these corner cases. At least two authors then regressively studied the effect of those event orders and verified that these tests were indeed flaky. We then reported these cases to the corresponding developers with the detailed reports on how to reproduce them. This report is initially necessary for validation, as these tests do not fail deterministically. 

\begin{table}[htb]
\caption{Common Exceptions and their frequencies}
\centering
\begin{adjustbox}{width=\linewidth}
	\small
 \begin{tabular}{|| l | l||}
 \hline
Reason & \# \\ [0.5ex]
 \hline
Waited for the root of the view hierarchy to have window focus & 97 \\
\hline
No views in hierarchy found matching & 36 \\
\hline
Could not inject Intent & 25 \\
\hline
Assertion Failed Error & 19 \\
\hline
Error performing single click: Animations are enabled on the target device & 15 \\
\hline
Unexpected state change & 12 \\
\hline
No such file or directory & 8 \\
\hline
java.lang.NoClassDefFoundError: android.support.test.espresso.intent.Intents & 6\\
\hline
TembaException: Unable to fetch ** & 6 \\
\hline
Attemp to invoke virtual method on a null reference object & 4 \\
\hline
MediaPlayer error & 4 \\
\hline
Wait for [**] to become idle timed out & 4 \\
\hline
others & 9 \\
\hline
\end{tabular}
\end{adjustbox}
\label{table:exception}
\end{table}

\subsubsection{Case Study: FirefoxLite-- SwitchSearchEngineTest}
\begin{figure}[t]
	\centering
	\includegraphics[width=\linewidth]{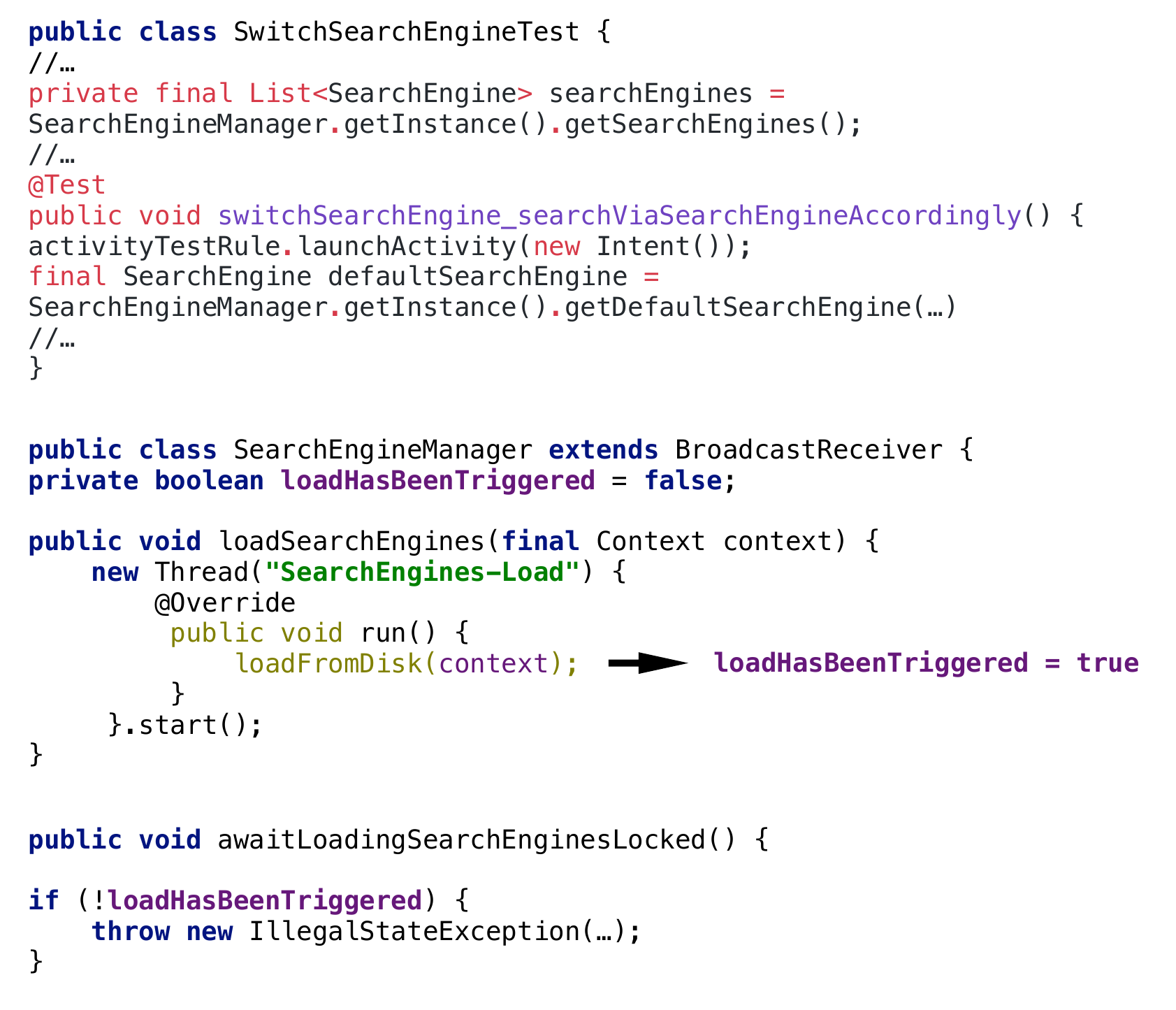}
		\caption{SwitchSearchEngineTest\label{listing:SwitchSearchEngineTest}
}
\end{figure}

%Figure~\ref{listing:SwitchSearchEngineTest} shows the parts of code snippet from the \emph{\wrapletters{SwitchSearchEngineTest}} test for the \emph{FirefoxLite} app. It tests the functionality provided by a broadcast receiver \emph{SearchEngineManager}, which initializes and loads different search engines. \emph{SwitchSearchEngineTest} starts its setup in line~\ref{lst:lineA_4} (Listing~\ref{listing:SwitchSearchEngineTest}) by loading the \emph{loadSearchEngines} method of \emph{SearchEngineManager} (Listing~\ref{listing:SearchEngineManager}, line~\ref{lst:lineB_4}). The \emph{loadSearchEngines} method creates a worker thread (Listing~\ref{listing:SearchEngineManager}, line~\ref{lst:lineB_11}) to initialize and load different search engines. After the setup, the test gets the default search engine (Listing~\ref{listing:SwitchSearchEngineTest}, line~\ref{lst:lineA_13}) by invoking the \emph{getDefaultSearchEngine} method of \emph{SearchEngineManager} (Listing~\ref{listing:SearchEngineManager}, line~\ref{lst:lineB_16}). This method checks if the loading of the search engine has been initiated, i.e., the worker thread has initialized the search engine. If the loading has not been initiated, it throws an \emph{IllegalStateException} (Listing~\ref{listing:SearchEngineManager}, line~\ref{lst:lineB_23}).
Figure~\ref{listing:SwitchSearchEngineTest} shows parts of code snippet from the \emph{\wrapletters{SwitchSearchEngineTest}} test for the \emph{FirefoxLite} app. It tests the functionality provided by a broadcast receiver \emph{SearchEngineManager}, which initializes and loads different search engines. During the startup process, search engines are loaded by \emph{loadSearchEngines} method. This method creates a worker thread (\emph{SearchEngines-Load}) to initiate the loading process in the background. Loading of search engines is verified (in main thread) via \emph{awaitLoadingSearchEnginesLocked}.

Evidently, we have three asynchronous threads: the espresso thread, the UIthread, and the worker thread. Worker threads in Android are implicitly moved into a background control group (\emph{cgroup}), where they only get a small percentage of the available CPU~\cite{threadscheduling}. In the scenario, where the worker thread has not started (waiting to get scheduled) and the espresso thread tries to access the default search engine, the above test fails. This depends on various factors such as the percentage of the CPU available and the number of background threads running. \tool detected the test as a flaky test by exploring different event execution orders. Since there is no synchronization between the testing thread and the worker thread, \tool delayed the worker thread at event schedule space identification phase such that the test failed and was reported the test as a flaky test.

%and  bIn event tracing, \tool executes statements one by one to verify that all events  are completely processed and that all threads have waited adequately for the results from async tasks. For this test, \tool delayed the worker thread (\emph{SearchEngines-Load}) by \emph{500 ms} to load the search engines, which is a reasonable amount of delay considering search engines are loaded from disk.
%%despite of the underline environment conditions.
%%The passing of the test depends on whether the worker thread has started and initialized the search engine loading. Detection of such flaky tests is difficult as the order and execution of threads depends on  and current approa , thus, the event, where the worker thread has not started and the espresso thread tries to access the default search engine, results in a failed test.
%

\section{Limitations}
\label{sec:discussion}
We identify the following potential limitations to our evaluation.
\begin{itemize}
\item {\em Event identification.} In our system, an event identifier is generated with data from two different threads. We may not successfully hook the event with its identifier for some rare cases, e.g., when our system runs extremely slowly, two pieces of data may not match the event identifier at the same time. Addressing this, we run our experiments on a system with a light workload and configure an emulator with a larger amount of memory (i.e., 8G). 
\item
{\em Thread dependency resolution.} In event schedule space identification, when too many threads are suspended at the same time, we terminate the execution and resolve thread dependencies to avoid imprecise schedule space identification.
\item
{\em Empirical study.} During our manual analysis on flaky tests, at least two of the authors analyze the log of a test failure for each flaky test to ensure the root cause of the test is correctly understood.
\end{itemize}
\section{Related Work}
\label{sec:relatework}

%Flaky test detection and fix
{\em Flaky test detection and fixing.} A few earlier researchers have started to work on flaky test issues. Bell et al.~\cite{deflaker} use code coverage differential analysis to identify flaky tests. A test is deemed flaky if it fails in the regression testing and its execution does not reach any code that was recently changed by developers.  Shi et al.~\cite{iFixFlakies} propose an approach to fix order-dependent flaky tests by leveraging passing tests.
Shi et al.~\cite{mitigating} propose to rerun a test multiple times on each mutant and obtain reliable coverage results such that the effects of flaky tests on mutation testing can be mitigated. Different from them, \tool detects concurrency-related flaky tests in Android apps by exploring feasible event execution orders.

{\em Event race detection.} Another branch of works that are close to ours is event race detection. Instead of detecting flaky tests, they leverage dynamic and static analysis to detect harmful event race. For instance, DROIDRACER~\cite{droidracer}, ERVA~\cite{erva}, EVENTRACER~\cite{scalable}, CAFA~\cite{race}, and nAdroid~\cite{nadroid} capture \emph{happens-before-relation} among events and inference possible event race errors. In addition, Ozkan et al.~\cite{asyncdroid} propose to detect asynchronous bugs by exploring different execution orders of event handlers in Android apps. These techniques have potential to apply to flaky test detetion, but face challenges to capture complete and precise happpen-before relations when a test is executed by a testing framework like Espresso. Many false positives can be reported due to incomplete happen-before relations as explained in Section~\ref{sec:rq2}. By contrast, \tool performs a system-level dynamic analysis to capture precise event dependencies to avoid such false positives.

%Flaky test studies
{\em Empirical studies on flaky tests.} Multiple studies~\cite{empirical, understanding, empiricalandroid} confirm concurrency as the major cause of flaky tests. Luo et al.~\cite{empirical} performed an empirical analysis of flaky tests in 51 open-source projects. They identified \emph{Concurrency} and \emph{Async wait} as the most common cause of flaky tests. They pointed out that the majority of these cases arose because they do not wait for external resources. Finally, they described the common fixing strategies the developers use to fix flaky tests. In a separate study, Eck et al.\cite{understanding} surveyed 21 professional developers to classify 200 flaky tests they fixed. They identified four unreported causes of flaky tests, which are also considered difficult to fix. Thorve et al.~\cite{empiricalandroid} conducted an empirical study of flaky tests in Android apps. They searched 1000 projects for the commits related to flakiness and found only 77 relevant commits from 29 projects. They found 36\% of commits occurred due to concurrency related issues. Fan et al.~\cite{apechecker} proposed a hybrid approach towards manifesting asynchronous bugs in Android apps. They studied 2097 apps and identified three async programming rules implied by the single-GUI-thread model. Based on these rules, they categorized three fault pattern and used static analysis to locate them in the app. Subsequently, they map these program traces to real event sequences to verify these errors.

%Testing strategies for concurrency
{\em Concurrency bugs detection.} There have been several testing based approaches~\cite{maple, nodefz, letko, covcon, contest, racket} to identify concurrency related bugs. Maple~\cite{maple} proposed a coverage-driven approach to expose untested thread interleavings. Letko~\cite{letko} proposed a combination of testing and dynamic analysis with \emph{metaheuuristic} techniques. Choudhary et al.~\cite{covcon} presented a coverage-guided approach for generating concurrency tests to detect bugs in thread-safe classes.
Multiple related works~\cite{dthreads, eventracecommander, delaybound, inputcovering, memoryunwinding, robustnesscon} manipulated event orders to control non-determinism in multi-threaded programs. Liu et al.~\cite{dthreads} proposed a deterministic multithreading system that replaces \emph{pthreads} library in C/C++ apps. Emmi et al.~\cite{delaybound} proposed a search prioritization strategy to discover concurrency bugs. They add non-determinism to deterministic schedulers by delaying their next-scheduled task. Adamsen et al.~\cite{eventracecommander} presented an automated program repair technique for event race errors in JavaScript. Given a repair policy, they controlled the event handler scheduling in the browser to avoid bad orderings.

\vspace{-1mm}
%Async bugs detection for Android

\section{conclusion}
\label{sec:conclusion}
Flaky tests pose a significant problem in validating mobile apps. Recent studies \cite{empirical, understanding, empiricalandroid} have shown concurrency as the most common cause of flaky tests. The uncertainty in a test outcome may arise due to synchronization issues originating from multiple threads interacting in a undesirable manner. In this paper, we presented an approach for detecting flaky tests through a systematic exploration of event orders. We introduced \tool, a tool to detect flaky tests for Android apps. \tool explores the space of all realizable execution environments where relevant threads interleave differently.  

Due to the lack of a testing benchmark for flaky tests, we created the first subject-suite \dataset that is used to study GUI test flakiness. \dataset contains 28 widely-used Android apps with 2.5k stars on average in GitHub. We applied \tool to tests from \dataset. Results show that \tool not only detected known flaky tests but also reported 245 new flaky tests. We believe that our tool and results hold out promise for the problem of tackling flaky tests, which is a significant pain point in industrial practice.
%The majority of the know Flaky tests in our dataset are due to non-deterministic execution environments.
%We demonstrated that our approach is efficient as compared to running a test multiple times.

%\balance
%\small
\bibliographystyle{ACM-Reference-Format}
\bibliography{main}

\end{document}